\documentclass[conference]{IEEEtran}

\usepackage{cite}
\usepackage{amsmath,amssymb,amsfonts}
\usepackage{algorithmic}
\usepackage{graphicx}
\usepackage{textcomp}
\usepackage[dvipsnames]{xcolor}
\usepackage{bm}

\definecolor{mypink}{RGB}{219, 48, 122}

\def\BibTeX{{\rm B\kern-.05em{\sc i\kern-.025em b}\kern-.08em
    T\kern-.1667em\lower.7ex\hbox{E}\kern-.125emX}}
    
\begin{document}

\title{Pain Detection with fNIRS-Measured Brain Signals: A Personalized Machine Learning Approach Using the Wavelet Transform and Bayesian Hierarchical Modeling with Dirichlet Process Priors
\thanks{Identify applicable funding agency here. If none, delete this.}
}
   
   \author{\IEEEauthorblockN{Daniel Lopez-Martinez$^{1,2}$,
   Ke Peng$^{3,4,5}$, Arielle Lee$^{3,4,6}$,
   David Borsook$^{3,4,6}$, 
   Rosalind Picard$^{2}$} \\
   \IEEEauthorblockA{$^1$Harvard-MIT Health Sciences and Technology, Massachusetts Institute of Technology \tt{dlmocdm@mit.edu}}
   \IEEEauthorblockA{$^2$Affective Computing group, MIT Media Lab, Massachusetts Institute of Technology}
   \IEEEauthorblockA{$^3$Center for Pain and the Brain, Harvard Medical School} 
   \IEEEauthorblockA{$^4$Department of Anesthesiology, Boston Children's Hospital}
   \IEEEauthorblockA{$^5$Neuo-epilepsy Lab, Centre de Recherche du CHUM, Montreal, Canada}
   \IEEEauthorblockA{$^6$Anthinoula A. Martinos Center for Biomedical Imaging, Department of Radiology, Massachusetts General Hospital}
   }


\maketitle

\begin{abstract}
Currently self-report pain ratings are the gold standard in clinical pain assessment. However, the development of objective automatic measures of pain could substantially aid pain diagnosis and therapy. Recent neuroimaging studies have shown the potential of functional near-infrared spectroscopy (fNIRS) for pain detection. This is a brain-imaging technique that provides non-invasive, long-term measurements of cortical hemoglobin concentration changes. In this study, we focused on fNIRS signals acquired exclusively from the prefrontal cortex, which can be accessed unobtrusively, and derived an algorithm for the detection of the presence of pain using Bayesian hierarchical modelling with wavelet features. This approach allows personalization of the inference process by accounting for inter-participant variability in pain responses. Our work highlights the importance of adopting a personalized approach and supports the use of fNIRS for pain assessment.

\end{abstract}


\section{Introduction}

Functional near infrared spectroscopy (fNIRS) is an emerging brain-imaging technique that provides non-invasive, long-term measurements of cortical hemoglobin concentration changes \cite{Obrig2014}. With fNIRS, near infrared light is projected onto the surface of the head, and is mainly absorbed by two types of hemoglobin during its propagation within the cerebral cortex, i.e. oxygenated hemoglobin (HbO) and deoxygenated hemoglobin (HbR). By continuously monitoring attenuation of light, fNIRS is able to reconstruct the cortical concentration changes of HbO and HbR, from which the local neuronal activity can be inferred. Compared with other imaging modalities such as functional magnetic resonance imaging (fMRI), fNIRS shares a similar physiological basis but has significant advantages in cost, robustness and portability, making it suitable to be used especially at bedside or in complex clinical settings. 

One novel application of fNIRS is to provide objective and robust assessment of pain. Pain is a subjective and complex experience that is processed within the human nervous system \cite{DeCWilliams2016}. While the current methods of assessing pain mostly rely on the participants’ self-report (the ``gold standard'') \cite{Twycross2015} or physiological signals (e.g. heart rate, blood pressure or skin conductance) \cite{Werner2014a,Werner2015a,dlmBHI2018, dlmNIPS2017, dlm_MTL_2017,Chu2017}, establishing an objective, robustness marker of pain in the central nervous system with a portable brain-imaging device would be beneficial for participants who cannot effectively communicate. In particular, patients undergoing surgery are unconscious during actual tissue damage. Without proper detection and control, repetitive pain signaling during surgery is postulated to lead to severe pain in the postoperative period and also potential initiation of chronic neuropathic pain \cite{Borsook2013a,Walker2015}.

Several previous studies employing fNIRS have revealed specific patterns in the cortical response to evoked pain stimuli from healthy volunteers, which were mainly characterized by activations (i.e. increases in HbO concentration and decreases in HbR concentrations, inferring elevated local neuronal activity) in the sensorimotor cortex (SMC) as well as deactivations (i.e. decreases in HbO concentration and increases in HbR concentration, inferring inhibited local neuronal activity)  in the medial prefrontal cortex (mPFC) \cite{Aasted2016,Yucel2015}. In our recent work exploring the possibility of combining fNIRS and machine learning to detect pain, we extracted features from both the SMC and the mPFC signals, and applied multi-task multiple kernel machine learning to account for the inter-participant variability in the brain response to pain \cite{dlm_ICPR_2018}. While this approach yielded moderate detection accuracy ($\sim$80\%), it is difficult to implement in the operating room, mainly because of the challenges of putting fNIRS optodes over the SMC in surgical patients who are usually kept in a supine position. Therefore, in this study, we decided to focus only on the prefrontal signals.

Previous work on pain detection has highlighted the importance of adopting a personalized machine learning approach to account for person-specific differences in pain responses \cite{dlmNIPS2017, dlm_MTL_2017, dlmCVPR2017, dlm_ICPR_2018}. Here, we used multi-task learning (MTL)  to learn personalized classifiers for the presence of pain while leveraging data from the entire population \cite{Caruana1997}. Specifically. we employed a non-parametric Bayesian hierarchical model to learn personalized logistic regression classifiers \cite{Xue2007} from time-frequency features derived using the continuous wavelet transform \cite{Cohen2003}. Our fNIRS-based machine learning approach was evaluated on a dataset containing pain responses to electrical noxious stimuli as well as data from baseline states.

The main contributions of this work are: (1) we present an approach to identify pain responses using fNIRS, (2) only fNIRS channels from the  prefrontal cortex were used, (3) we show that by personalizing the machine learning models using multi-task learning the system is able to provide differentiation of pain state with better performance. This approach could have great promise for pain assessment in noncommunicative patients as well as multiple clinical populations, and form the basis of a more objective measure of pain than  self-report.

\section{Data}

\begin{figure}
	\centering
	\includegraphics[width=0.41\linewidth]{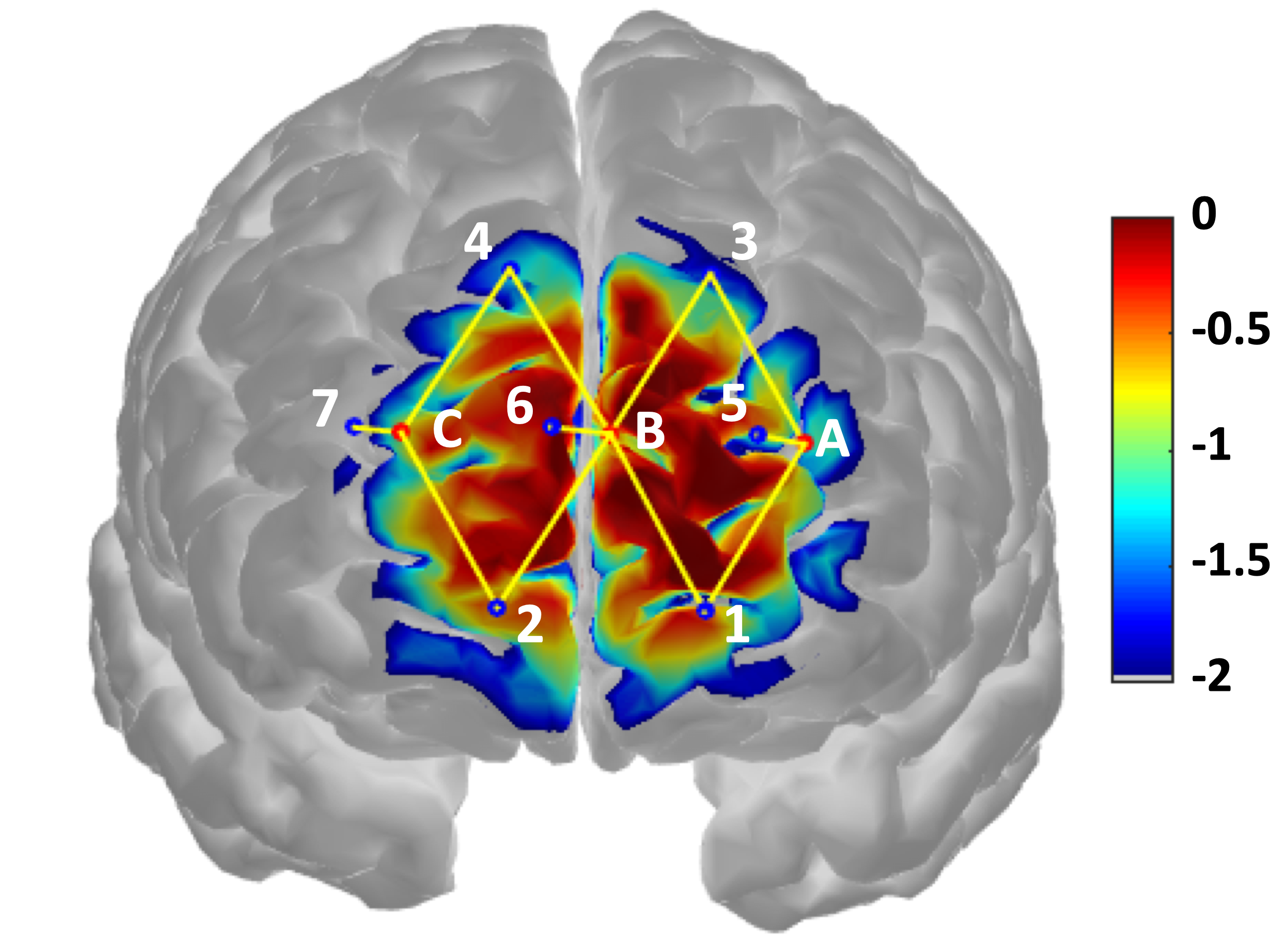}
	\caption{Frontal view of the brain cortex showing the arrangement of fNIRS optodes with the corresponding sensitivity profile in mm$^{-1}$. The sources (A, B, C) and detectors (1-7) cover the brain's prefrontal cortex and result in 8 channels, shown as yellow lines.}
	\label{fig:mount}
\end{figure}

\subsection{Data acquisition}

This study was approved by the institutional review boards of Massachusetts General Hospital and Boston Children's Hospital. The datasets of a total of 43 healthy  participants (age range: 26.8 $\pm$ 5.6, 20 to 39 years old) were included in this study. Previous work has shown sex differences in pain-related cerebral processing and blood oxygenation signals \cite{Riley1998,Moulton2006}. Hence, in this work we decided to focus on a single gender  to control the number of varying factors in the data. Because pain-related brain activation in females is also be influenced by menstrual cycles \cite{Veldhuijzen13}, only male participants were enrolled. Written consent was obtained from each of the participating participants prior to the study procedures.

A TechEn CW7 continuous wave fNIRS system (TechEn Inc., Massachusetts, USA) was used for data acquisition. Three light emitters and 4 light detectors were mounted to cover the prefrontal areas of the brain, forming 8 emitter-detector pairs (called fNIRS channels; see Fig.\ref{fig:mount}) with inter-optode distance of 3 cm to enable detection sensitivity of cortical hemodynamic activities. Two wavelengths of near infrared light were used: 690 nm and 830 nm. One intrinsic limitation of fNIRS is that the recorded data contains not only the hemodynamic signals from the cerebral cortex but also the contributions from several extracerebral layers (e.g. skin, scalp and skull) because near infrared light has to penetrate those layers before and after its travel within the cortex. Those extracerebral signals are mostly related to the non-neuronal physiological effects of the body including heartbeat, respiration and blood pressure. To correct for such contamination, we installed a short distance detector 8 mm away from each of the light emitters to just sample the hemodynamic changes from those extracerebral layers. The recorded short distance data were later used in the general linear model (GLM) analysis as regressors (see below).

During the data acquisition period, each participant underwent the following three scanning sessions supervised by study personnel: a resting-state session, a tactile brush session and an electrical stimulation session. During the resting state, the participant was asked to simply sit comfortably in a chair and rest for 6 min. In the tactile brush session, a study personnel delivered 12 non-painful brush stimuli by gently brushing the dorsum of the participant’s left hand. Each brush stimulus lasted for 5 sec and was followed by a resting period of 25 sec. In the electrical stimulation session, two levels of 5 Hz electrical pulses were applied by the study personnel to the participant’s left thumb with a neurometer (Neurotron, Maryland, USA): an innocuous level (score: 3/10) for which the participant described that he was strongly aware but did not feel any pain, as a noxious level (score: 7/10) for which the participant reported much pain but was able to tolerate without breath holding, sweating or any retreat actions. The intensities of the two electrical stimulation levels were determined with a few pre-tests before the actual scanning session. Similar to the tactile brush stimulus, each electrical stimulus lasted 5 sec and was separated from one another with a 25 sec resting period. The entire electrical stimulation sequence contained a total of 6 innocuous stimuli and 6 noxious stimuli. The order of the innocuous and noxious stimuli was randomized for each participant to accommodate for the effect of pain anticipation.

\subsection{Data preprocessing} \label{sec:datapreprocessing}

The acquired fNIRS data were first processed with an open-source toolbox HomER2 \cite{homer} based on Matlab (Mathworks, Massachusetts, USA). 
The measured light intensity of the  channels (see Fig.\ref{fig:mount}) was then converted to optical density changes by taking the logarithm of the signal. 
Motion artifacts were then detected and corrected with the hybrid Spline Interpolation and Savitzky-Golay filtering method \cite{Jahani2018}. A temporal low-pass filter with a cutoff frequency of 0.5 Hz was applied to remove high frequency oscillations in the data that presumably did not have a neuronal basis (e.g. heart beat whose frequency is around 1 Hz). The processed optical density timecourses were then transformed to the concentration changes of HbO and HbR using the modified Beer-Lambert Law \cite{Delpy1988,Cope1988,Boas2004} with a partial pathlength factor of 6. The concentration changes of total hemoglobin (HbT) were obtained by a direct summation of the HbO and HbR changes (HbT = HbO + HbR). For each of the tactile brush and electrical stimulation sessions, we then reconstructed the hemodynamic response function (HRF) to tactile brush or to noxious and innocuous electrical stimuli from 2 sec before the onset of the stimulus to 20s after the onset for each of the three types of hemoglobin using a GLM \cite{Peng2018}. In the model, the HRFs to stimuli were approximated with consecutive Gaussian basis functions with width and temporal inter-spacing to be both 1 sec. For each normal fNIRS channel, the short separation measurement that had the highest correlation was also included in the GLM as a regressor to filter out the physiological signal from extracerebral layers. Polynomial terms up to the 3rd order were also added to the model to remove the low frequency drifts from the signal.

\subsection{Feature extraction} \label{sec:features}

\begin{figure}
	\centering
	\includegraphics[width=0.75\linewidth]{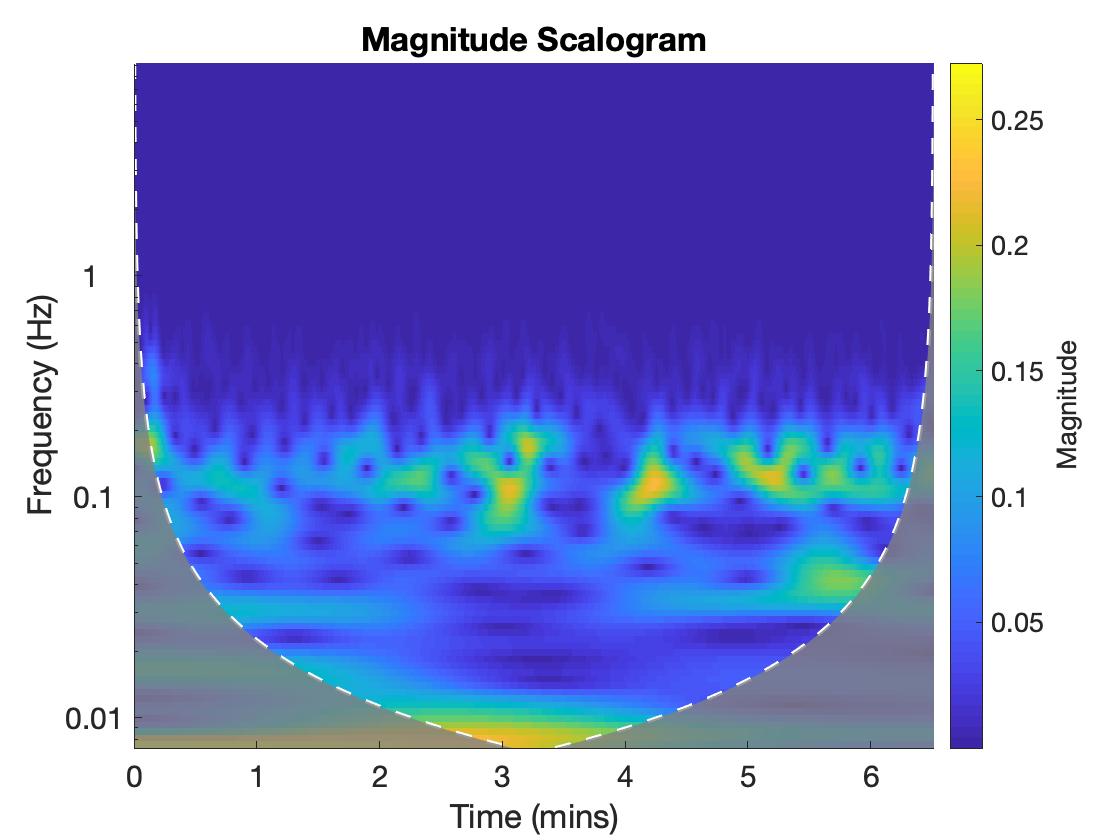}
	\caption{Exemplary time-frequency analysis of a processed HbO signal (see Sec.\ref{sec:datapreprocessing}) with the Morlet wavelet transform.}
	\label{fig:timefreqanalysis}
\end{figure}

Following \cite{dlm_ICPR_2018}, we extracted windows of duration 20 seconds from the HbO signals. These windows were either "pain"  or "no-pain" windows. Pain windows were extracted starting immediately after the onset of the  electrical noxious stimulus (score: 7/10) and were assigned label $y=1$. No-pain windows were randomly sampled from the pre-stimulation baseline (resting) recording and were assigned label $y=0$.

Feature extraction was performed using the discretized continuous wavelet transform \cite{Cohen2003}, a powerful time-frequency analysis technique that provides a  description of the power spectrum of the signal in terms of both time and frequency domains simultaneously. Hence, it remedies the drawbacks of traditional time-domain feature extraction (inability to characterize the frequency components of the fNIRS signals) and frequency-domain feature extraction  (temporal dynamics are lost).

Wavelets are functions with zero mean that are both time and frequency  localized. In this work, we use the Morlet Wavelet, which is given by:
\begin{equation}
\Psi_{w_{o}}(t)=c_{w_{o}} \pi^{-\frac{1}{4}} e^{-\frac{1}{2} t^{2}}\left(e^{i w_{o} t}-e^{-\frac{1}{2} w_{o}^{2}} \right)
\end{equation}
where $ c_{w_{o}}=\left(1+e^{-w_{o}^{2}}-2 e^{-\frac{3}{4} w_{o}^{2}}\right)^{-\frac{1}{2}} $ is a normalization constant, $w_{0}$ is the wavelet central frequency and $t$ is time \cite{Addison2002}. The wavelet transform is defined as follows:
\begin{equation}
T(a, b)=\frac{1}{\sqrt{a}} \int_{-\infty}^{+\infty} x(t) \psi^{*}\left(\frac{t-b}{a}\right) d t
\end{equation}
where $x(t)$ is the signal analyzed, $a$ is the dilation of the wavelet, $b$ is the location of the wavelet and $\psi^{*}(t)$ is the complex conjugate of the wavelet function $\psi(t)$. This continuous wavelet transform was discretized using ten voices per octave.

Following the wavelet transformation, we extracted the scalogram (see Fig.\ref{fig:timefreqanalysis} for an example), which represents the strength or energy of each coefficient, using the $L_2$ norm. From the scalogram, we  extracted features from two frequency bans:  very-low frequency oscillations (VLFO, 0.01-0.08 Hz) and low frequency oscillations (LFO, 0.08-0.15 Hz). Specifically, we extracted the following features in the wavelet domain: (1,2,3) the mean, maximum and standard deviation of $|T(a,b)|$ in $a=[0.01,0.08]$, (4,5,6) the mean, maximum and standard deviation of $|T(a,b)|$ in $a=[0.08,0.15]$, (7,8) the location of the maximum and slope of $\int_{a=0.01}^{0.08}|T(a,b)| da / \int_{a=0.01}^{0.08}da $, (9,10) the location of the maximum and slope of $\int_{a=0.08}^{0.15}|T(a,b)| da / \int_{a=0.08}^{0.15}da $.
This was done for each of the 8 fNIRS channels depicted in Fig.\ref{fig:mount}, resulting in a feature vector of dimension $D=80$. All features were then normalized prior to being used in our ML models described in Sec.\ref{sec:pml}.

\begin{figure}
	\centering
	\includegraphics[width=0.6\linewidth]{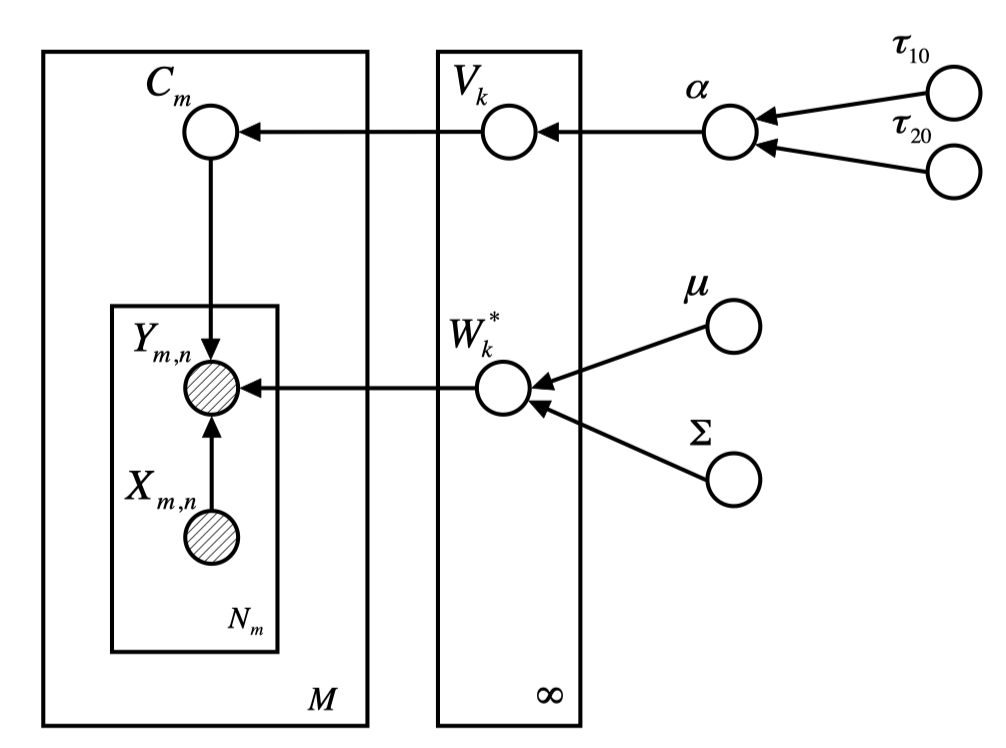}
	\caption{Graphial representation of the hierarchical Bayesian MTL model \cite{Xue2007}.}
	\label{fig:model}
\end{figure}

\section{Personalized machine learning model} \label{sec:pml}

In this work, we consider the binary classification task of detecting the presence of pain using multi-task learning. Traditional single-task learning (STL) refers to the approach of learning a classifier for a single classification task using the corresponding dataset $\mathcal{D} = \{ (\bm{x}_i, y_i )\}_{i=1}^{N}$, where $\bm{x}_i \in \mathbb{R}^D$ is a $D$-dimensional feature vector, $y_i \in \{ 0,1 \}$ is a binary label and $N$ is the number of samples in the dataset. Many classification tasks can be viewed as consisting of multiple correlated subtasks, such as pain detection in a particular patient subpopulation that responds to noxious stimuli in a characteristic way, similar to yet different from other patient subpopulations \cite{dlm_ICPR_2018, dlmNIPS2017, dlm_MTL_2017, dlmCVPR2017}. Therefore, pooling all subtasks and treating them as a single task may not be appropriate. Furthermore, the traditional STL approach of isolating each subtask and learning the corresponding classifier independently may not be appropriate neither, as it does not exploit the potential information that may be acquired from other correlated classification subtasks. 


Multi-task learning (MTL) is a type of inductive transfer learning in which multiple classifiers are learned simultaneously for several related tasks using a shared representation \cite{Caruana1997}.
This  is distinct from traditional single-task learning in several aspects: (1) multiple tasks are learned in parallel, which may represent e.g. individuals or groups of individuals that share certain characteristics, (2) these tasks are not identical, so that pooling them together into a  single task is not appropriate, (3) because tasks are correlated, what is learned form one task may be transferable to another, (4) by learning tasks in parallel, tasks can exploit the potential information one may acquire from other classiﬁcation tasks, hence leveraging the limited amount of training data available for each task, to the benefit of all.
Furthermore, exploiting data from related tasks leads to better generalization of the resulting models. 

In this work, following previous work on multi-task learning for predicting mood, stress, and health \cite{Taylor2017}, we used hierarchical bayesian logistic regression (HBLR). This a nonparametric hierarchical Bayesian MTL model with a common prior  drawn from the Dirichlet process (DP) that learns logistic regression classifiers. This model, introduced in \cite{Xue2007}, results in personalized logistic regression classifiers that account for individual characteristics in brain hemodynamic responses to pain, while learning from the entire population. Moreover, our approach induces a soft clustering of the tasks without any \textit{a priori} knowledge or meta-information. This may be used to uncover novel pain phenotypes.

Let's consider $M$ tasks with corresponding datasets $\mathcal{D}_m = \{ (\bm{x}_{m,i}, y_{m,i} )\}_{i=1}^{N_m}$, where $m \in \{1, ..., M \}$ is the task and $N_m$ is the number of instances for task $m$.
Following \cite{Xue2007,Taylor2017}, we model the conditional distribution of $y_{m,i}$ given $\bm{x}_{m,i}$ via logistic regression:

{ \footnotesize
\begin{equation}
p\left(y_{m, i} | \bm{w}_{m}, \bm{x}_{m, i}\right)=\sigma\left(\bm{w}_{m}^{T} \bm{x}_{m, i}\right)^{y_{m, i}}\left[1-\sigma\left(\bm{w}_{m}^{T} \bm{x}_{m, i}\right)\right]^{1-y_{m, i}}
\end{equation} }
where $\sigma(x) = \frac{1}{1+\exp{(-x)}}$ and $\bm{w}_m$ represents the classifier weights for task $m$. The goal is to learn $\{ \bm{w}_m \}_{m=1}^{M}$ jointly, sharing information between tasks. 
To do so, the hierarchy consists of a bottom layer with task-specific parameters, and a top layer in which tasks are connected together via a common prior. Given the prior, individual models are learned independently. However, the common prior is also learned during training, resulting in information transfer between tasks.

As in \cite{Xue2007}, the common prior $G$ on the task-specific model parameters in the proposed nonparametric hierarchical Bayesian model is drawn from the Dirichlet process (DP):
\begin{equation}
\begin{aligned} w_{m} | G & \sim G \\ G & \sim D P\left(\alpha, G_{0}\right) \end{aligned}
\end{equation}
where $\alpha \sim G a\left(\tau_{10}, \tau_{20}\right)$ is the positive innovation parameter drawn from a Gamma distribution and $G_{0} \sim N_{d}(\bm{\mu} =\bm{0}, \bm{\Sigma}=\sigma \bm{I})$ is $d$-dimensional multivariate normal base distribution. The DP prior is employed due to its implicit non-parametric soft-clustering mechanism, which enables the model to  automatically identify the similarities between the various tasks and adjust the complexity of the model, that is, the number of task clusters. This also provides  valuable insights into tasks that exhibit similar behaviours. Because the clustering is \textit{soft}, tasks need not be assigned discretely to single clusters. Instead, tasks can belong to many clusters in varying degrees. The innovation parameter $\alpha >0$ controls the probability of creating a new cluster,  with larger $\alpha$ yielding more clusters. When $\alpha \rightarrow \infty$ there is a cluster for each task, whereas small values of $\alpha$ lead to only a few distinct clusters. In this work, $\alpha$ has the following probability distribution function  in the shape-rate parametrization:
\begin{equation}
p\left(\alpha | \tau_{10}, \tau_{20}\right)=\frac{\tau_{20}^{\tau_{10}}}{\Gamma\left(\tau_{10}\right)} \alpha^{\tau_{10}-1} \exp \left(-\tau_{20} \alpha\right)
\end{equation}
where $\Gamma(\cdot)$ is the Gamma function. Integrating $\alpha$ over a diffuse hyper-prior, as opposed to defining it directly, increases the roboustness of the algorithm \cite{Xue2007}.

In Bayesian modeling, we are interested in the posterior distribution of the latent variables given the training data and hyperparamters:
{ \small \begin{equation}
p\left(\mathbf{Z} |\left\{\mathcal{D}_{m}\right\}_{m=1}^{M}, \bm{\Phi}\right)=\frac{p\left(\left\{\mathcal{D}_{m}\right\}_{m=1}^{M} | \mathbf{Z}, \mathbf{\Phi}\right) p(\mathbf{Z} | \mathbf{\Phi})}{p\left(\left\{\mathcal{D}_{m}\right\}_{m=1}^{M} | \Phi\right)}
\end{equation}
}
where $\bm{Z} =
\{\{\bm{c}_{m}\}_{m=1}^{M},\{v_{k}\}_{k=1}^{\infty}, \alpha,\{\bm{w}_{k}^{*}\}_{k=1}^{\infty}\}$ denotes the collection of latent variables, $\bm{\Phi}=\{\tau_{10}, \tau_{20}, \bm{\mu}, \bm{\Sigma}\}$ denotes the collection of given parameters and hyper-parameters and $\bm{c}_m$ is an all-zero vector except that the $k$-th entry is equal to one if task $m$ belongs to cluster $k$.
To approximate $p\left(\mathbf{Z} |\left\{\mathcal{D}_{m}\right\}_{m=1}^{M}, \Phi\right)$, whose computation is intractable, we used mean-field variational bayesian (VB) inference \cite{Ghahramani2001}. This approximates the true posterior by a variational distribution $q(\bm{Z})$ and converts computation of posteriors into an optimization problem. Following \cite{Blei2006}, we adopted a truncated stick-breaking representation for the variational distribution. By setting the truncation level equal to the number of desired clusters $K$, we can control the number of resulting clusters and the complexity of the HBLR model. The factorized variational distribution is then specified as:
{ 
\begin{equation} {\scriptscriptstyle
q(\mathbf{Z})=\left[\prod_{m=1}^{M} q_{c_{m}}\left(c_{m}\right)\right] \cdot\left[\prod_{k=1}^{K} q_{v_{k}}\left(v_{k}\right)\right] \cdot q_{\alpha}(\alpha) \cdot\left[\left(\prod_{k=1}^{K} q_{w_{k}^{*}}\left(w_{k}^{*}\right)\right]\right.
} \end{equation}
}
where $c_{m} \sim M_{K}\left(1 ; \phi_{m, 1}, \ldots, \phi_{m, K}\right), m=1, \ldots, M$, $v_{k} \sim B e\left(\varphi_{1, k}, \varphi_{2, k}\right), k=1, \ldots, K-1$ and $
w_{k}^{*} \sim N_{d}\left(\theta_{k}, \Gamma_{k}\right), k=1, \ldots, K$ (see \cite{Xue2007} for details).

To learn all the parameters, we use a coordinate ascent algorithm  \cite{Xue2007} that uses the mean-field approach \cite{Ghahramani2001}. All hype-parameters were initialized following \cite{Taylor2017}, and re-estimated iteratively until convergence.

The prediction function for a new test sample $x_{m,*}$ becomes:

{ \small
\begin{equation}
\begin{aligned} 
p(y_{m,*}=1 | x_{m,*}, \Phi,&\left\{\theta_{k}\right\}_{k=1}^{K},\left\{\Gamma_{k}\right\}_{k=1}^{K} ) \\
=& \sum_{k=1}^{K} \phi_{k}^{(t)} \int \sigma\left(w_{k}^{* T} x_{m,*}\right) N_{d}\left(\theta_{k}, \Gamma_{k}\right) d w_{k}^{*} \\
\approx & \sum_{k=1}^{K} \phi_{k}^{(t)} \sigma\left(\frac{\theta_{k}^{T} x_{m,*}}{\sqrt{1+\frac{\pi}{8} x_{m,*}^{T} \Gamma_{k} x_{m,*}}}\right)
\end{aligned}
\end{equation}
}
where we use the approximate form of the integral \cite{MacKay1992} and $\sigma (x) = 1/(1+ e^{-x})$ is the sigmoid function.

\begin{table}[]
\centering
\caption{Classification performance of single-task models}
\label{tab:stml}
\begin{tabular}{lrrrr}
\hline
\textbf{Model}           & \textbf{Accuracy} & \textbf{Pr.} & \textbf{R.} & \textbf{F$_1$} \\ \hline
Logistic regression (L1) & 0.58                & 0.59            & 0.55           & 0.55             \\
Logistic regression (L2) & 0.60               & 0.61            & 0.57           & 0.58             \\
SVM (linear kernel)      & 0.57               &  0.58            & 0.55           & 0.55             \\
SVM (rbf kernel)         & \textbf{0.69}                & 0.72            & 0.63           & 0.67            
\end{tabular}
\end{table}
\begin{table}[]
\centering
\caption{Classification performance of the HBLR model}
\label{tab:hblr}
\begin{tabular}{lrrrr}
\hline
\textbf{K} & \textbf{Accuracy} & \textbf{Pr.} & \textbf{R.} & \textbf{F$_1$} \\ \hline
2   & 0.66              & 0.53   &    0.53    &   0.53\\
3   & 0.73              & 0.67   &    0.53    &   0.59\\
4   & \textbf{0.81}     & 0.75   &    0.71    &   0.73\\
5   & 0.75              & 0.67   &    0.59    &   0.63\\
          
\end{tabular}
\end{table}



\section{Results}

After processing the fNIRS signals as described in Sec.\ref{sec:datapreprocessing}, windows of duration 20 seconds were extracted from the HbO signals. The choice of signal modality (that is, HbO) and window size was informed by previous work done on pain detection from fNIRS \cite{Pourshoghi2016,dlm_ICPR_2018}. From these windows, we extracted the $D=80$ features described in Sec.\ref{sec:features} from the prefrontal fNIRS channels depicted in Fig.\ref{fig:mount}. All features were normalized, and the dataset was balanced by downsampling the over-represented class (no pain; $y=0$), resulting in $N=510$ instances from $M=43$ tasks with both classes equally represented.

First, we evaluated traditional single-task machine learning models using 10-fold cross-validation. Namely, we used logistic regression with $L_1$ and $L_2$ regularization \cite{dlm_regularization_2017}, and support vector machines with linear and radial basis function (RBF) kernels. The later resulted in the highest accuracy (0.69) among these models, as shown in Table \ref{tab:stml}.

Following the single-task models, we shifted to the multi-task setting 
and evaluated the multi-task HBLR model with different values of $K$, which corresponds to different truncation levels or number of clusters. To do so, for each of the 43 tasks we randomly split the data into training (90\%) and test partitions (10\%) using 10-fold cross-validation. The results are shown in Table \ref{tab:hblr} and indicate improved performance of the HBLR model with respect to single-task models. The best performance was achieved with $K=4$, and significantly outperforms the single-task models. We also evaluated $K>5$ (not shown in Table \ref{tab:hblr}), but the additional number of clusters did not result in improved performance. For every value of $K$, model hyperparameters were optimized using 10-fold cross-validation. The best performing model used $\tau_{10}=0.01$, $\tau_{20}=0.1$.

\begin{figure}
	\centering
	\includegraphics[width=0.985\linewidth]{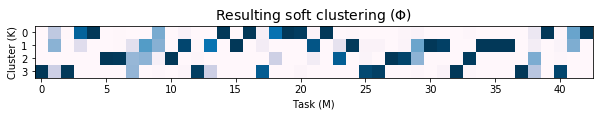}
	\caption{Soft clustering of $M=43$ tasks into $K=4$ clusters by the HBLR model, with darker cells indicating a higher degree of membership into the cluster.}
	\label{fig:softclustering}
\end{figure}

The HBLR model performs a soft clustering of the $M=43$ tasks into $K$ clusters, such that a given task may belong to multiple clusters with varying degrees of membership. This is visualized in Fig.\ref{fig:softclustering} for the best performing model with $K=4$. While the analysis of the clustering results may be used to uncover novel pain phenotypes, this was hampered in this work  by the limited dataset size, low heterogeneity in the participant pool, as well as lack of meta-information.

\section{Discussion and Conclusion}

This study investigated the use of functional near-infrared spectroscopy (fNIRS) for the detection of evoked pain, using fNIRS signals derived from optodes placed exclusively on the frontal brain cortex. Multiple studies have pointed towards an integrated role of the prefrontal cortex in the high-level process of pain perception \cite{dlm_ICPR_2018,Peng2017}. Whereas previous studies (e.g. \cite{dlm_ICPR_2018}) have included other brain regions such as the sensorimotor cortex, focusing only on prefrontal cortical area has many advantages including its ease of access (especially in surgical settings where the patient is kept in a supine position) and the ability to avoid hair contamination in the recorded fNIRS signals. Moreover, the need of only sampling from the prefrontal cortex  greatly reduces the required number of fNIRS optodes and optical fibres to be mounted on the participant's head, leading to an easier installation process and more patient comfort. Hence, this work investigated the use of frontal fNIRS channels exclusively. 
One central challenge in establishing objective and reliable assessment of pain with brain signals is the notable inter-subject variability in pain perception and responses, which hampers the ability of machine learning models to generalize across people. Previous work has shown improved pain assessment performances with personalized machine learning models \cite{dlmNIPS2017, dlm_MTL_2017, dlmCVPR2017, dlm_ICPR_2018}. Hence, this work employed multi-task machine learning to provide personalized analysis. 



We evaluated our model on a dataset containing fNIRS responses to experimental electrical pain. Our results confirmed previous work \cite{dlm_ICPR_2018} that showed that fNIRS-measured brain signals and machine learning may be used to detect the presence of pain. Using exclusively frontal channels, we showed that high pain state discrimination accuracy may be achieved. Furthermore, our results confirmed the importance of adopting a multi-task approach to personalize the assessment.

This work has several limitations. The strength of the model is limited by the size of our dataset (43 participants) and lack of heterogeneity in the participant pool of healthy males. Moreover, only the detection of experimental evoked electrical pain at a single pain intensity was evaluated. Future work should extend our model to females, other participant phenotypes, different pain stimuli, varying pain intensities, and other types of pain (e.g. ongoing or chronic pain). Also, the model was evaluated using pain windows  aligned with the onset of the stimulus.  Future work should evaluate the model continuously (see \cite{dlmEMBC2018} for an example).  Despite the limitations, this novel fNIRS personalized machine-learning approach to pain detection showed great promise in its accuracy, using only prefrontal electrodes.  Thus, this work advances progress toward objective pain recognition, providing accurate and practical measures of pain that work even if patients lack alertness or the ability to articulate their experience. It may also help advance the development of automatic analgesia administration systems for hospital settings that currently rely on subjective self-reported pain measures \cite{Lopez-Martinez2019}. 

{\small
\bibliographystyle{IEEEtran}
\bibliography{Mendeley}

\begin{thebibliography}{10}
\providecommand{\url}[1]{#1}
\csname url@samestyle\endcsname
\providecommand{\newblock}{\relax}
\providecommand{\bibinfo}[2]{#2}
\providecommand{\BIBentrySTDinterwordspacing}{\spaceskip=0pt\relax}
\providecommand{\BIBentryALTinterwordstretchfactor}{4}
\providecommand{\BIBentryALTinterwordspacing}{\spaceskip=\fontdimen2\font plus
\BIBentryALTinterwordstretchfactor\fontdimen3\font minus
  \fontdimen4\font\relax}
\providecommand{\BIBforeignlanguage}[2]{{%
\expandafter\ifx\csname l@#1\endcsname\relax
\typeout{** WARNING: IEEEtran.bst: No hyphenation pattern has been}%
\typeout{** loaded for the language `#1'. Using the pattern for}%
\typeout{** the default language instead.}%
\else
\language=\csname l@#1\endcsname
\fi
#2}}
\providecommand{\BIBdecl}{\relax}
\BIBdecl

\bibitem{Obrig2014}
\BIBentryALTinterwordspacing
H.~Obrig, ``{NIRS in clinical neurology - a 'promising' tool?}''
  \emph{NeuroImage}, vol.~85, pp. 535--546, 1 2014.
\BIBentrySTDinterwordspacing

\bibitem{DeCWilliams2016}
A.~C. d.~C. Williams and K.~D. Craig, ``{Updating the definition of pain},''
  \emph{PAIN}, vol. 157, no.~11, pp. 2420--2423, 11 2016.

\bibitem{Twycross2015}
A.~Twycross, T.~Voepel-Lewis, C.~Vincent, L.~S. Franck, and C.~L. von Baeyer,
  ``{A Debate on the Proposition that Self-report is the Gold Standard in
  Assessment of Pediatric Pain Intensity},'' \emph{The Clinical Journal of
  Pain}, vol.~31, no.~8, pp. 707--712, 8 2015.

\bibitem{Werner2014a}
P.~Werner, A.~Al-Hamadi, R.~Niese, S.~Walter, S.~Gruss, and H.~C. Traue,
  ``{Automatic Pain Recognition from Video and Biomedical Signals},'' in
  \emph{2014 22nd International Conference on Pattern Recognition}.\hskip 1em
  plus 0.5em minus 0.4em\relax IEEE, 8 2014, pp. 4582--4587.

\bibitem{Werner2015a}
S.~Gruss, R.~Treister, P.~Werner, H.~C. Traue, S.~Crawcour, A.~Andrade, and
  S.~Walter, ``{Pain Intensity Recognition Rates via Biopotential Feature
  Patterns with Support Vector Machines},'' \emph{PLOS ONE}, vol.~10, no.~10,
  10 2015.

\bibitem{dlmBHI2018}
D.~Lopez-Martinez and R.~Picard, ``{Skin conductance deconvolution for pain
  estimation},'' in \emph{IEEE Conference on Biomedical and Health Informatics
  (BHI)}, Las Vegas, 2018.

\bibitem{dlmNIPS2017}
D.~Lopez-Martinez, O.~Rudovic, and R.~Picard, ``{Physiological and Behavioral
  Profiling for Nociceptive Pain Estimation Using Personalized Multitask
  Learning},'' in \emph{Neural Information Processing Systems (NIPS) Workshop
  on Machine Learning for Health}, Long Beach, USA, 2017.

\bibitem{dlm_MTL_2017}
\BIBentryALTinterwordspacing
D.~Lopez-Martinez and R.~Picard, ``{Multi-task neural networks for personalized
  pain recognition from physiological signals},'' in \emph{2017 Seventh
  International Conference on Affective Computing and Intelligent Interaction
  Workshops and Demos (ACIIW)}, San Antonio, TX, 10 2017, pp. 181--184.
\BIBentrySTDinterwordspacing

\bibitem{Chu2017}
\BIBentryALTinterwordspacing
Y.~Chu, X.~Zhao, J.~Han, and Y.~Su, ``{Physiological Signal-Based Method for
  Measurement of Pain Intensity},'' \emph{Frontiers in Neuroscience}, vol.~11,
  no. May, pp. 1--13, 5 2017.
\BIBentrySTDinterwordspacing

\bibitem{Borsook2013a}
\BIBentryALTinterwordspacing
D.~Borsook, B.~D. Kussman, E.~George, L.~R. Becerra, and D.~W. Burke,
  ``{Surgically Induced Neuropathic Pain},'' \emph{Annals of Surgery}, vol.
  257, no.~3, pp. 403--412, 3 2013.
\BIBentrySTDinterwordspacing

\bibitem{Walker2015}
\BIBentryALTinterwordspacing
S.~M. Walker, ``{Pain after surgery in children},'' \emph{Current Opinion in
  Anaesthesiology}, vol.~28, no.~5, pp. 570--576, 10 2015.
\BIBentrySTDinterwordspacing

\bibitem{Aasted2016}
C.~M. Aasted, M.~A. Yucel, S.~C. Steele, K.~Peng, D.~A. Boas, L.~Becerra, and
  D.~Borsook, ``{Frontal lobe hemodynamic responses to painful stimulation: A
  potential brain marker of nociception},'' \emph{PLoS ONE}, vol.~11, no.~11,
  pp. 1--12, 2016.

\bibitem{Yucel2015}
M.~A. Y{\"{u}}cel, C.~M. Aasted, M.~P. Petkov, D.~Borsook, D.~A. Boas, and
  L.~Becerra, ``{Specificity of hemodynamic brain responses to painful stimuli:
  a functional near-infrared spectroscopy study.}'' \emph{Scientific reports},
  vol.~5, p. 9469, 2015.

\bibitem{dlm_ICPR_2018}
\BIBentryALTinterwordspacing
D.~Lopez-Martinez, K.~Peng, S.~C. Steele, A.~J. Lee, D.~Borsook, and R.~Picard,
  ``{Multi-task multiple kernel machines for personalized pain recognition from
  functional near-infrared spectroscopy brain signals},'' in \emph{2018 24th
  International Conference on Pattern Recognition (ICPR)}.\hskip 1em plus 0.5em
  minus 0.4em\relax Beijing: IEEE, 8 2018, pp. 2320--2325.
\BIBentrySTDinterwordspacing

\bibitem{dlmCVPR2017}
\BIBentryALTinterwordspacing
D.~Lopez-Martinez, O.~Rudovic, and R.~Picard, ``{Personalized Automatic
  Estimation of Self-Reported Pain Intensity from Facial Expressions},'' in
  \emph{2017 IEEE Conference on Computer Vision and Pattern Recognition
  Workshops (CVPRW)}.\hskip 1em plus 0.5em minus 0.4em\relax IEEE, 7 2017, pp.
  2318--2327.
\BIBentrySTDinterwordspacing

\bibitem{Caruana1997}
R.~Caruana, ``{Multitask Learning},'' \emph{Machine Learning}, vol.~28, no.~1,
  pp. 41--75, 1997.

\bibitem{Xue2007}
Y.~Xue, X.~Liao, L.~Carin, B.~Krishnapuram, and B.~K. Com, ``{Multi-Task
  Learning for Classification with Dirichlet Process Priors},'' \emph{Journal
  of Machine Learning Research}, vol.~8, pp. 35--63, 2007.

\bibitem{Cohen2003}
\BIBentryALTinterwordspacing
L.~Cohen, \emph{{Wavelets and Signal Processing}}, L.~Debnath, Ed.\hskip 1em
  plus 0.5em minus 0.4em\relax Boston, MA: Birkh{\"{a}}user Boston, 2003.
\BIBentrySTDinterwordspacing

\bibitem{Riley1998}
\BIBentryALTinterwordspacing
J.~L. Riley, M.~E. Robinson, E.~A. Wise, C.~D. Myers, and R.~B. Fillingim,
  ``{Sex differences in the perception of noxious experimental stimuli: a
  meta-analysis.}'' \emph{Pain}, vol.~74, no. 2-3, pp. 181--7, 2 1998.
\BIBentrySTDinterwordspacing

\bibitem{Moulton2006}
\BIBentryALTinterwordspacing
E.~A. Moulton, M.~L. Keaser, R.~P. Gullapalli, R.~Maitra, and J.~D. Greenspan,
  ``{Sex differences in the cerebral BOLD signal response to painful heat
  stimuli},'' \emph{American Journal of Physiology-Regulatory, Integrative and
  Comparative Physiology}, vol. 291, no.~2, pp. R257--R267, 8 2006.
\BIBentrySTDinterwordspacing

\bibitem{Veldhuijzen13}
\BIBentryALTinterwordspacing
D.~S. Veldhuijzen, M.~L. Keaser, D.~S. Traub, J.~Zhuo, R.~P. Gullapalli, and
  J.~D. Greenspan, ``{The role of circulating sex hormones in menstrual
  cycle–dependent modulation of pain-related brain activation},''
  \emph{Pain}, vol. 154, no.~4, pp. 548--559, 4 2013.
\BIBentrySTDinterwordspacing

\bibitem{homer}
\BIBentryALTinterwordspacing
T.~J. Huppert, S.~G. Diamond, M.~A. Franceschini, and D.~A. Boas, ``{HomER: a
  review of time-series analysis methods for near-infrared spectroscopy of the
  brain.}'' \emph{Applied optics}, vol.~48, no.~10, pp. 280--98, 4 2009.
\BIBentrySTDinterwordspacing

\bibitem{Jahani2018}
\BIBentryALTinterwordspacing
S.~Jahani, S.~K. Setarehdan, D.~A. Boas, and M.~A. Y{\"{u}}cel, ``{Motion
  artifact detection and correction in functional near-infrared spectroscopy: a
  new hybrid method based on spline interpolation method and Savitzky–Golay
  filtering},'' \emph{Neurophotonics}, vol.~5, no.~01, p.~1, 2 2018.
\BIBentrySTDinterwordspacing

\bibitem{Delpy1988}
\BIBentryALTinterwordspacing
D.~T. Delpy, M.~Cope, P.~V.~D. Zee, S.~Arridge, S.~Wray, and J.~Wyatt,
  ``{Estimation of optical pathlength through tissue from direct time of flight
  measurement},'' \emph{Physics in Medicine and Biology}, vol.~33, no.~12, pp.
  1433--1442, 12 1988.
\BIBentrySTDinterwordspacing

\bibitem{Cope1988}
M.~Cope and D.~T. Delpy, ``{System for long-term measurement of cerebral blood
  and tissue oxygenation on newborn infants by near infra-red
  transillumination.}'' \emph{Medical {\&} biological engineering {\&}
  computing}, vol.~26, no.~3, pp. 289--94, 5 1988.

\bibitem{Boas2004}
D.~A. Boas, A.~M. Dale, and M.~A. Franceschini, ``{Diffuse optical imaging of
  brain activation: Approaches to optimizing image sensitivity, resolution, and
  accuracy},'' \emph{NeuroImage}, vol.~23, no. SUPPL. 1, 2004.

\bibitem{Peng2018}
K.~Peng, M.~A. Y{\"{u}}cel, C.~M. Aasted, S.~C. Steele, D.~A. Boas, D.~Borsook,
  and L.~Becerra, ``{Using prerecorded hemodynamic response functions in
  detecting prefrontal pain response: a functional near-infrared spectroscopy
  study.}'' \emph{Neurophotonics}, vol.~5, no.~1, 1 2018.

\bibitem{Addison2002}
\BIBentryALTinterwordspacing
P.~Addison, J.~Watson, and T.~Feng, ``{Low-Oscillation Complex Wavelets},''
  \emph{Journal of Sound and Vibration}, vol. 254, no.~4, pp. 733--762, 7 2002.
\BIBentrySTDinterwordspacing

\bibitem{Taylor2017}
S.~A. Taylor, N.~Jaques, E.~Nosakhare, A.~Sano, and R.~Picard, ``{Personalized
  Multitask Learning for Predicting Tomorrow's Mood, Stress, and Health},''
  \emph{IEEE Transactions on Affective Computing}, 2017.

\bibitem{Ghahramani2001}
\BIBentryALTinterwordspacing
Z.~Ghahramani and M.~J. Beal, ``{Propagation Algorithms for Variational
  Bayesian Learning},'' in \emph{Advances in Neural Information Processing
  Systems 13}, 2001, pp. 507--513.
\BIBentrySTDinterwordspacing

\bibitem{Blei2006}
D.~M. Blei and M.~I. Jordan, ``{Variational inference for Dirichlet process
  mixtures},'' \emph{Bayesian Analysis}, vol.~1, no.~1, pp. 121--144, 2006.

\bibitem{MacKay1992}
\BIBentryALTinterwordspacing
D.~J.~C. MacKay, ``{The Evidence Framework Applied to Classification
  Networks},'' \emph{Neural Computation}, vol.~4, no.~5, pp. 720--736, 9 1992.
\BIBentrySTDinterwordspacing

\bibitem{Pourshoghi2016}
\BIBentryALTinterwordspacing
A.~Pourshoghi, I.~Zakeri, and K.~Pourrezaei, ``{Application of functional data
  analysis in classification and clustering of functional near-infrared
  spectroscopy signal in response to noxious stimuli},'' \emph{Journal of
  Biomedical Optics}, vol.~21, no.~10, 2016.
\BIBentrySTDinterwordspacing

\bibitem{dlm_regularization_2017}
\BIBentryALTinterwordspacing
D.~Lopez-Martinez, ``{Regularization approaches for support vector machines
  with applications to biomedical data},'' \emph{arXiv}, 10 2017.
\BIBentrySTDinterwordspacing

\bibitem{Peng2017}
\BIBentryALTinterwordspacing
K.~Peng, S.~C. Steele, L.~Becerra, and D.~Borsook, ``{Brodmann area 10:
  Collating, integrating and high level processing of nociception and pain},''
  \emph{Progress in Neurobiology}, 2017.
\BIBentrySTDinterwordspacing

\bibitem{dlmEMBC2018}
\BIBentryALTinterwordspacing
D.~Lopez-Martinez and R.~Picard, ``{Continuous Pain Intensity Estimation from
  Autonomic Signals with Recurrent Neural Networks},'' in \emph{Engineering in
  Medicine and Biology Society (EMBC) Conference}.\hskip 1em plus 0.5em minus
  0.4em\relax Hawaii: IEEE, 7 2018.
\BIBentrySTDinterwordspacing

\bibitem{Lopez-Martinez2019}
D.~Lopez-Martinez, P.~Eschenfeldt, S.~Ostvar, M.~Ingram, C.~Hur, and R.~Picard,
  ``{Deep Reinforcement Learning for Optimal Critical Care Pain Management with
  Morphine using Dueling Double-Deep Q Networks},'' in \emph{Engineering in
  Medicine and Biology Society (EMBC) Conference}, 2019.

\end{thebibliography}
}

\end{document}